\documentclass{ws-p9-75x6-50}
    
\def\be{\begin{equation}}
\def\ee{\end{equation}}
\def\bea{\begin{eqnarray}}
\def\eea{\end{eqnarray}}
\def\lbl{\label}

\begin{document}
		
\title{On junction conditions in gravity theories with higher curvature
        terms}
\author{Tom\'a\v s Dole\v zel}
 \address {D\'epartement d'Astrophysique Relativiste et de Cosmologie,\\ 
UMR 8629 du Centre National de la Recherche Scientifique,\\ 
Observatoire de Paris, 92195 Meudon, France}

\address{Institute of Theoretical Physics, Charles University,\\ 
V Hole\v sovi\v ck\'ach 2, 18000 Prague 8, Czech Republic}
\date{\today}

\maketitle

\abstracts{We discuss the junction conditions in the context of the Randall-Sundrum model with the Gauss-Bonnet interaction.
We consider the $Z_2$ symmetric model where the brane is embedded in an $AdS_5$ bulk, 
as well as a model without $Z_2$ symmetry in which the brane (in this case called by tradition ``shell'') separates two
metrically different $AdS_5$ regions. 
We show that the Israel junction conditions across the membrane (that is either a brane or a shell)
have to be modified if more general equations than Einstein's,
including higher curvature terms, hold in the bulk, as is likely to be the case in a low energy limit of string theory. We find that the
membrane can then no longer be treated in the thin wall
approximation. We derive the junction conditions for the
Einstein-Gauss-Bonnet theory including second order curvature terms
and show that the microphysics of Gauss-Bonnet thick membranes may, in some
instances, be simply hidden in a renormalization of Einstein's
constant.}

\section*{I Introduction}

Motivated by recent developments in high energy physics [1,2,3,4] there is at present a considerable
increase of activity in the domain of cosmology with extra dimensions. 
In these models
gravity is assumed to act in a $n$-dimensional ``bulk'' while the standard
model interactions are confined to a 4-dimensional slice (``brane'' worldsheet) of this multi-dimensional
spacetime. Randall and Sundrum
have recently proposed two models in which all the matter  
is confined to a 4-dimensional brane worldsheet embedded in a 5-dimensional 
anti-de Sitter ($AdS_5$) spacetime with imposed $Z_2$ metric symmetry, $w\to-w$ ($w$ denotes the fifth dimension and the
metric is expressed in the Gaussian normal coordinates) [5].

What is usually done in General Relativity, when, e.g., studying the gravitational
collapse
of spherical bodies is to join two {\it metrically different} solutions of the Einstein field equations.  
Integration of the Einstein 
equations across the surface separating the two regions leads to the Israel junction conditions ([6], see
also Appendix A for a review) relating the surface stress-energy tensor to the discontinuity of the extrinsic curvature
across the surface. The sign of the extrinsic curvature, and thus the form of the junction
conditions, depend on the definitions of normal vectors in the neighbourhood of the surface. However, once the
directions of the normal vectors on each side of the surface are fixed, e.g. pointing in a defined positive sense, 
the formalism becomes unambiguous and yields in general
a nonzero stress-energy tensor for the surface (massive ``shell''). 

On the other hand matter of $Z_2$ symmetric $AdS_5$ branes, which connect two {\it metrically identical} solutions of the Einstein equations, arises
formally by a flip of the normal vectors at one or the other side of the boundary surface while
preserving the form of Israel junction conditions.  
This flip of the normals across the brane 
describes the connection, via the Israel junction conditions, of a bulk region 
and its mirror image. Were no formal flip of the normal vectors performed, i.e. dropping out of the $Z_2$ symmetry,
one would join in a topologically trivial way 
two complementary parts of the $AdS_5$ spacetime across a non-massive boundary surface.
Hence, the $Z_2$ symmetry dropped out, a massive shell must separate two $AdS_5$ 
regions with different cosmological constants [7]. 

If these brane or shell models are to be the low energy limit of string theory, it is likely that the field
equations include higher curvature terms. In particular, the lowest non-linear curvature terms derive from   
the Gauss-Bonnet Lagrangian, which, in five dimensions, is the only
non-linear term in the curvature which yields second order field equations (see e.g. [8] and references therein).
The Randall-Sundrum model with the Gauss-Bonnet correction has recently been considered in ref. [9-11].
However, only in [7] are the peculiarities of the junction conditions
carefully examined.
 
In this paper based on [7] we derive the junction conditions across the membrane
(either brane or shell) taking into account the Gauss-Bonnet correction and
conclude that in this case the thin wall approximation fails.
When second (and higher) curvature terms are taken into account the junction conditions formally include products of 
distributions as well as products of distributions and non-infinitely differentiable functions: 
they are therefore not well defined in the distributional sence.
The complete evolution of the Einstein-Gauss-Bonnet 
universe must be studied more carefully taking into account the internal structure of the membrane.
Nevertheless, the microphysics of the thick membrane may, at late times, be hidden in a renormalization of Einstein's constant.

\section*{II Einstein-Gauss-Bonnet membranes}

The membrane cosmological models with the $AdS_5$ bulk are solutions of the Einstein equations
$G_{AB}+\Lambda g_{AB}=0$ everywhere except on the membrane $\Sigma$. In brane cosmology the cosmological
constant
$\Lambda $ is the same on each side of $\Sigma$, in shell cosmology it jumps from $\Lambda_+$ to
$\Lambda_-$. Now, in order to take into account the Gauss-Bonnet correction, we shall consider the gravitational action
\be S_g=\int\! d^5\! x\,\sqrt{-g}(-2\Lambda +R+\alpha L_2)\lbl{4.1}\ee
with
\be L_2=R_{ABCD}R^{ABCD}-4R_{AB}R^{AB}+R^2\ee
where $\alpha$ is a coupling constant, and where $R_{ABCD}$,  $R_{AB}$ and $R$ are the Riemann tensor, the Ricci tensor
and the scalar curvature of the five dimensional metric $g_{AB}$ with determinant $g$.  The corresponding field
equations, outside the membrane, are 
\be\Lambda g_{AB}+G_{AB}+\alpha H_{AB}=0\lbl{4.2}\ee
with
\be H_{AB}\equiv 2R_{ALMN}R_B^{\ \ LMN}-4R_{AMBN}R^{MN}-4R_{AM}R_B^{\ \ M}+2RR_{AB}-{1\over2}g_{AB}L_2.\ee

Contrarily to Einstein's equations, the equations (\ref{4.2})  possess, for $\alpha\neq0$ and a given value of the cosmological
constant $\Lambda $, two (anti) de Sitter solutions 
\bea R_{ABCD}&=&L_\pm(g_{AC}g_{BD}-g_{AD}g_{BC})\nonumber\\
\hbox{with}\qquad L_\pm&=&{1\over4\alpha}\left(-1\pm\sqrt{1+{4\alpha\Lambda\over3}}\right).\lbl{4.3}\eea
In brane cosmology we
shall choose one or the other solution everywhere in the bulk. In shell cosmology, we shall choose the
solution
$L_+$ on one side of the shell and the solution $L_-$ on the other side. (Shell cosmologies are therefore more
satisfactory in Einstein-Gauss-Bonnet theory as one does not have to impose different cosmological constants on each
side of the shell.)

In order to get the stress-energy tensor on the membrane, one proceeds along Israel's line.
We first choose a Gaussian coordinate system $(w,x^\mu)$ such that the metric reads
\bea ds^2&=&dw^2+\gamma_{\mu\nu}dx^\mu dx^\nu\nonumber\\
&=&dw^2-n^2(\tau,w)d\tau^2+S^2(\tau,w)[d\chi^2+f^2_k(\chi)(d\theta^2+\sin^2\theta
d\phi^2)],\lbl{4.4}\eea 
$w=0$ being the equation of the membrane $\Sigma$
and functions  $n(\tau,w)$, $S(\tau,w)$ being given in ref. [12].
In this coordinate system  the extrinsic curvature of the surfaces $w=constant$ is simply given by
\be K_{\mu\nu}=-{1\over2}{\partial\gamma_{\mu\nu}\over\partial w}.\lbl{4.5}\ee
It jumps across the membrane from $K^+_{\mu\nu}$ to $K^-_{\mu\nu}$ (with $K^+_{\mu\nu}=-K^-_{\mu\nu}$ in the
case of branes) and this discontinuity can be described in terms of the Heaviside distribution.

Expressing now the Riemann tensor (\ref{4.3}) in terms of $K_{\mu\nu}$ and the four dimensional Riemann tensor of the
metric $\gamma_{\mu\nu}$  we then obtain from (\ref{4.2}), everywhere outside the membrane (see Appendices A and B for
the '4+1' decompositions of $G^A_{~B}$ and $H^A_{~B}$)
\be\Lambda \delta_{~\mu}^\nu+G_{~\mu}^\nu+\alpha H_{~\mu}^\nu= (1+4\alpha L)
\left({\partial K_{~\mu}^\nu\over\partial w}-\delta_{~\mu}^\nu{\partial
K\over\partial w}\right)+...\qquad(=0)\lbl{4.6}\ee
 where $K\equiv
\gamma^{\alpha\beta}K_{\alpha\beta}$, where $L=L_+$ or $L_-$, and  where the dots stand for terms containing at most
first order
$w$-derivatives of
$\gamma_{\mu\nu}$.

In Einstein's theory, $\alpha=0$, and (\ref{4.6}), in the vicinity of $\Sigma$, is well defined in a distributional sense:
$\partial K_{~\mu}^\nu/\partial w$ can be expressed in terms of the Dirac distribution and the integration of (\ref{4.6}) across
the membrane gives Israel's junction conditions, that is the stress-energy tensor on the membrane in terms of the
jump in the extrinsic curvature, eq. (\ref{A4}). 

When $\alpha\neq0$ on the other hand, (\ref{4.6}) is not well defined in a distributional sense, as $L$ cannot be considered as
an infinitely $w$-differentiable function. Indeed,  in shell cosmology, $L$ jumps from $L_+$ to $L_-$ across
$\Sigma$, and in brane cosmology, $L=L_+=L_-$ is continuous across $\Sigma$, but, because of the reflexion
symmetry, has a discontinuous $w$-derivative. This mathematical obstruction simply means that, in
Einstein-Gauss-Bonnet theory, membranes cannot be treated in the thin wall approximation: the jumps in the extrinsic
curvature and in $L$ or its derivative have to be described in detail within  specific microphysical models. 

When the thickness of the  membrane is taken into account, the distributions $\partial K_{~\mu}^\nu/\partial w$ and $L$ 
are replaced by rapidly varying but $C^\infty$ functions. Supposing that the metric keeps the form (\ref{4.4}), we can
define from the
$\tau$-$\tau$ component of (\ref{4.6}) the sharply peaked function (cf. (\ref{BH}) in Appendix B) 
\be\kappa\rho\equiv 3(1+4\alpha L){\partial K_{~\chi}^\chi\over\partial w}\lbl{4.7}\ee
(The $\chi$-$\chi$ component of (\ref{4.6}) is redundant thanks to the conservation
equation, that is the Bach-Lanczos identity (\ref{BI}).)
In the vicinity and inside the membrane $K_{~\chi}^\chi$ can be written as
\be K_{~\chi}^\chi={1\over2}\bar K_{~\chi}^\chi+{1\over2}\hat K_{~\chi}^\chi \,f(\tau,w)\lbl{4.8}\ee
where $\bar K_{\chi\chi}\equiv K^+_{\chi\chi}+K^-_{\chi\chi}$ and  where the function $f(\tau,w)$, which  varies
rapidly from $-1$ to $+1$ across $\Sigma$, encapsulates its microphysics. Similarly, in the case of brane cosmology, we
can write
\be L=\tilde Lg_b(\tau,w)\lbl{4.9}\ee
where $\tilde L=L_+$ or $L_-$ and where $g_b(\tau,w)$ is some even function of
$w$ which varies rapidly from $+1$ to $+1$ across the brane. In shell cosmology on the other hand
\be L={1\over2}\bar L+{1\over2}
\hat L\,g_s(\tau,w)\lbl{4.10}\ee
where $g_s(\tau,w)$ varies rapidly from $-1$ to $+1$ and $\bar L=-{1\over2\alpha}$, $\hat L={1\over2\alpha}\sqrt{1+
{4\alpha\Lambda\over3}}$.

Integrating (\ref{4.7}) across $\Sigma$ we therefore get the energy density of the membrane as
\be\kappa\varrho\equiv\int_{-\eta}^{+\eta}\!dw\,\kappa\rho 
=3\hat K^\chi_{~\chi}\left[1+2\alpha\tilde L\int_{-\eta}^{+\eta}\!dw \,
g_b{\partial f\over\partial w}\right]\lbl{4.11}\ee
in the case of branes, and
\be\kappa\varrho\equiv\int_{-\eta}^{+\eta}\!dw\,\kappa\rho 
={3\over2}\sqrt{1+{4\alpha\Lambda\over3}}\,\hat
K^\chi_{\chi}\int_{-\eta}^{+\eta}\!dw \, g_s{\partial f\over\partial
w}\lbl{4.12}\ee
 in the case of shells (the fact that one does not recover the results of Einstein's theory when $\alpha=0$ is not
surprising as $L_{\pm}$ is divergent in that case) . Now, if $f$ or $g_{b/s}$ depend on $\tau$, the integrals in 
(\ref{4.11})-(\ref{4.12}) are some functions of
$\tau$. But, if $f$ and $g_{b/s}$ do not depend on time, which is probably to be expected whence the brane has reached a
stationary state, that is at late times, then the integrals in (\ref{4.11})-(\ref{4.12}) 
are just  numbers. In this case then, the
microphysics of the membrane is simply hidden in a renormalization of the Einstein constant $\kappa$.
 
The decomposition of $H^A_{~B}$ is given in full generality in Appendix B, eq. (\ref{BBB}). Higher curvature terms typically induce
terms with higher powers in extrinsic curvature, or products of components of the Riemann and extrinsic curvature 
tensors. Since these terms are obviously no well defined in the distributional sense, the thin wall formalism is no longer applicable beyond the linear
order in curvature tensors. The behaviour of the membranes in the context of such theories needs to be studied more carefully, 
taking into account the microphysics of the thick membrane, as is done in e.g. [16].   
Nevertheless, at late times, the microphysics of an Einstein-Gauss-Bonnet membrane can be hidden in a renormalization of Einstein's constant. 



\appendix

\section{Junction conditions for non-null surfaces in General Relativity}
This appendix summarizes the junction conditions in the theory of Einstein 
(Lanczos [13], Darmois [14], Misner and Sharp [15], Israel [6]). 
Suppose we are given a 4-dimensional hypersurface ($\Sigma$) in a 5-dimensional spacetime (metric $g_{AB}$) 
which can be imagined as the element of a family of surfaces. The normal vectors $n^A$ to this 
family of surfaces are not null; $n_A n^A\equiv \epsilon=\pm 1$. They are all oriented in a
positive direction defined in the bulk. 
Let the surface be either spacelike ($\epsilon=-1$) or timelike ($\epsilon=+1$). 
As an aid in deriving junction conditions we introduce Gaussian normal coordinates 
in the neighbourhood of $\Sigma$. The metric $g_{AB}$ has the form
\be ds^2=\epsilon dw^2+\gamma_{\mu\nu}dx^{\mu}dx^{\nu},\lbl{A1}\ee         
and the extrinsic curvature of the surfaces $w=constant$ is 
\be K_{\mu\nu}=-{1\over 2}{\partial\gamma_{\mu\nu}\over\partial w}.\ee
The curvature tensor of the metric $g_{AB}$ can be expressed in terms of the intrinsic curvature 
of 4-dimensional hypersurface (metric $\gamma_{\mu\nu}$) and of its extrinsic curvature;
one gets the so-called Gauss-Codazzi equations. 
In the special case of Gaussian normal coordinates the equations simplify to
\bea
R_{w\mu w\nu}&=&{\partial K_{\mu\nu}\over \partial w}+K_{\rho\nu}K^{\rho}_{\,\,\mu}\,,\lbl{R1}\\
R_{w\mu\nu\rho}&=&\nabla_{\nu}K_{\mu\rho}-\nabla_{\rho}K_{\mu\nu}\,,\\
R_{\lambda\mu\nu\rho}&=&~^4R_{\lambda\mu\nu\rho}+\epsilon\left[K_{\mu\nu}K_{\lambda\rho}-
                                                                K_{\mu\rho}K_{\lambda\nu}\right]\lbl{R3}\,,
\eea 
where $\nabla_{\rho}$ is the covariant derivative with respect to the 4-dimensional metric $\gamma_{\mu\nu}$. 
From (\ref{R1})-(\ref{R3}) we obtain the decomposition
of the Ricci tensor ($R_{AB}=g^{CD}R_{CADB}$) and of the scalar curvature ($R=g^{AB}R_{AB}$) as:
\bea
R_{ww}&=&\gamma^{\mu\nu}{\partial K_{\mu\nu}\over \partial w}+Tr(K^2)\,,\\
R_{w\mu}&=&\nabla_{\mu}K-\nabla_{\nu}K^{\nu}_{\,\,\mu}\,,\\
R_{\mu\nu}&=&~^4R_{\mu\nu}+\epsilon\left[{\partial K_{\mu\nu}\over\partial w}+2K_{\mu}^{\,\,\rho}K_{\rho\nu}
                                                 -K K_{\mu\nu} \right]\,,\\
R&=&~^4R+\epsilon\left[2\gamma^{\mu\nu}{\partial K_{\mu\nu}\over\partial w}+3Tr(K^2)-K^2\right]\lbl{RK}\,,  
\eea
where we defined $K\equiv K^{\mu}_{\,\,\mu}$ and $Tr(K^2)\equiv K^{\mu\nu}K_{\mu\nu}$. 

In terms of the intrinsic and extrinsic curvature of the 4-dimensional hypersurfaces $w=constant$, the 
Einstein tensor ($G_A^{~B}=R_A^{~B}-(1/2)\delta_A^{~B}R\,$) and the field equations have components
\bea
G^w_{~w}&=&-{1\over 2}~^4R+{1\over 2}\epsilon\left[K^2-Tr(K^2)\right]=\kappa T^w_{~w}\,,\lbl{G1}\\
G^w_{~\mu}&=&\epsilon\left[\nabla_{\mu}K-\nabla_{\nu}K^{\nu}_{\,\,\mu}\right]=\kappa T^w_{~\mu}\,,\lbl{G2}\\
G^{\mu}_{~\nu}&=&~^4G^{\mu}_{~\nu}
              +\epsilon\left[{\partial K^{\mu}_{~\nu}\over\partial w}-\delta^{\mu}_{~\nu}
                   {\partial K\over\partial w}\right]\nonumber\\
              &&+\epsilon\left[- 
                   K K^{\mu}_{~\nu}+{1\over 2}\delta^{\mu}_{~\nu}Tr(K^2)+{1\over 2}\delta^{\mu}_{~\nu} K^2\right]
             =\kappa T^{\mu}_{~\nu}\,.\lbl{G3}
\eea 
If the stress-energy tensor $T^A_{~B}$ contains a 'delta-function contribution' at $\Sigma$, the integral
of $T^A_{~B}$ with respect to the proper distance $w$ measured perpendicularly through $\Sigma$,
\be {\cal T}^A_{~B}\equiv\lim_{\eta\to 0}\left[\int_{-\eta}^{\eta}T^A_{~B}dw\right],\lbl{A3}\ee
is non-zero and represents the surface stress-energy tensor. In this case the extrinsic curvature must be
a distribution of 'Heaviside type' at $\Sigma$. Integral (\ref{A3}) applied on equations (\ref{G1})-(\ref{G3}) yields the 
junction conditions relating the stress-energy tensor of $\Sigma$ to the discontinuity of the extrinsic curvature
at $\Sigma$. In the passage to the limit $\eta\to 0$ only the terms $\sim (\partial K^{\mu}_{~\nu}/\partial w)$ contribute
to yield 
\bea
\kappa {\cal T}^w_{~w}&=&0\,,\nonumber\\
\kappa {\cal T}^w_{~\mu}&=&0\,,\nonumber\\                      
\kappa {\cal T}^{\mu}_{~\nu}&=&\epsilon(\hat K^{\mu}_{~\nu} -\delta^{\mu}_{~\nu}\hat K)\,,
\quad\hbox{where}\quad \hat K^{\mu}_{~\nu}\equiv K^{\mu}_{~\nu}(0+)-K^{\mu}_{~\nu}(0-)\lbl{A4}.
\eea
It is useful to denote $K^+_{\mu\nu}\equiv K_{\mu\nu}(0+)$, $K^-_{\mu\nu}\equiv K_{\mu\nu}(0-)$.

As for the intrinsic geometry of $\Sigma$, it must be continuous across $\Sigma$; this is the second junction contition
completing equations (\ref{A4}). If there are no 'delta singularities' contained in $T^A_{~B}$, the bulk is sliced by
massless 'boundary surfaces'.

\section{Junction conditions for the theory of Einstein-Gauss-Bonnet}
The theory of Einstein-Gauss-Bonnet is based on the following Lagrangian\footnote{We analyse the particular case of 
a 5-dimensional spacetime sliced by 4-dimensional hypersurfaces.}
\be {\cal L}=\sqrt{-g}\left[-2\Lambda+R+\alpha L_2\right],\ee 
where $g$ is the determinant of the 5-dimensional metric $g_{AB}$ and $\alpha$ is a constant of the dimension 
of $[length]^2$.
$L_2$ is the Gauss-Bonnet Lagrangian which reads
\be L_2=R_{ABCD}R^{ABCD}-4R_{AB}R^{AB}+R^2.\ee
The Euler variation of ${\cal L}$ gives the following field equations:
\be \Lambda g_{AB}+G_{AB}+\alpha H_{AB}=0.\ee
$G_{AB}$ is the Einstein tensor and $H_{AB}$ is its analogue stemmed from the Gauss-Bonnet part of the
Lagrangian, $L_2$,
\bea 
G_{AB}&\equiv& R_{AB}-{1\over2}g_{AB}R\,,\\
H_{AB}&\equiv& 2\left[R_{ALMN}R_B^{\,\,\,LMN}-2R_{AMBN}R^{MN}-2R_{AM}R_B^{\,\,\,M}+RR_{AB}\right]\nonumber\\
             &&-{1\over 2} g_{AB}L_2.\lbl{GB}\quad
\eea
$G_{~B}^A$ and $H_{~B}^A$ satisfy the Bianchi and Bach-Lanczos identities respectively
\be \nabla_A G_{~B}^A=0\quad,\quad\nabla_A H_{~B}^A=0\lbl{BI}.\ee

In order to derive the junction condition in the theory of Einstein-Gauss-Bonnet 
we need to express $H_{AB}$ in terms of the intrinsic curvature of hypersurfaces $w=constant$ and their
extrinsic curvatures. We adopt the notation used in Appendix A in which     
the '4+1' decomposition of the Einstein tensor $G_{AB}$ is shown (expressions (\ref{G1})-(\ref{G3})).

Inserting the decomposition of the curvature tensors (\ref{R1})-(\ref{RK}) into (\ref{GB}) one finds the following 
results: $H^A_{~B}$ does not contain terms $\sim (\partial K^{\mu}_{~\nu}/\partial w)^2$ as one would expect since
$H^A_{~B}$ contains terms $\sim (R_{ABCD})^2$. Further, there are no terms linear in 
($\partial K^{\mu}_{~\nu}/\partial w$) in $H^w_{~w}$ and  $H^w_{~\mu}$ so that these junction 
conditions correspond to those in the theory of Einstein: ${\cal T}^w_{~w}={\cal T}^w_{~\mu}=0$.
The components $H^{\mu}_{~\nu}$ are
\bea
H^{\mu}_{~\nu}&=&\left\{{\partial K^{\mu}_{~\nu}\over \partial w}\right\}\left(2Tr(K^2)-2K^2\right)
                  +\left\{{\partial K^{\mu}_{~\lambda}\over \partial w}\right\}\left(4KK^{\lambda}_{\,\,\nu}
                                                              -4K_{\nu\beta}K^{\beta\lambda}\right)\nonumber\\
                 &+&\left\{{\partial K_{\nu\lambda}\over \partial w}\right\}\left(4KK^{\lambda\mu}
                                                              -4K^{\mu}_{~\beta}K^{\beta\lambda}\right)
                  +\left\{{\partial K \over \partial w}\right\}\left(4K^{\mu}_{~\beta}K^{\beta}_{\,\,\nu}
                                                              -4KK^{\mu}_{~\nu}\right)\nonumber\\
                 &+&\left\{{\partial K_{\alpha\beta}\over \partial w}\right\}\left(4K^{\mu}_{~\nu}K^{\alpha\beta}
                                                              -4K^{\alpha\mu}K^{\beta}_{\,\,\nu}\right)\nonumber\\
                 &+&\left\{{\partial K_{\alpha\beta}\over \partial w}\right\}\delta^{\mu}_{~\nu}
                                                                                   \left(4K^{\alpha}_{\,\,\gamma}
                                                       K^{\gamma\beta}-4KK^{\alpha\beta}\right)
                 +\left\{{\partial K \over \partial w}\right\}\delta^{\mu}_{~\nu}\left(2K^2-2Tr(K^2)\right)\nonumber\\
                 &+&\epsilon\left(-4~^4R_{~\alpha\nu}^{\mu~~~\beta}{\partial K^{\alpha}_{~\beta}\over\partial w}
                                  -4~^4R^{\alpha}_{~\nu}{\partial K^{\mu}_{~\alpha}\over\partial w}
                                  -4~^4R^{\alpha\mu}{\partial K_{\nu\alpha}\over\partial w}\right)\nonumber\\ 
                 &+&\epsilon\left(4~^4R^{\mu}_{~\nu}{\partial K \over\partial w}
                   +2~^4R{\partial K^{\mu}_{~\nu}\over\partial w}
                   +4\delta^{\mu}_{~\nu}~^4R^{\alpha\beta}{\partial K_{\alpha\beta}\over\partial w}                                               
  -2\delta^{\mu}_{~\nu}~^4R{\partial K \over\partial w}  \right)\nonumber\\
                 &+&\ldots\,\lbl{BBB}, 
\eea   
where '\ldots' includes terms of zeroth order in $(\partial K^{\mu}_{~\nu}/\partial w)$ which disappear in the passage
to the limit $\eta\to 0$ of the integration (\ref{A3}).

In the case the metric has the form
$$ds^2=dw^2-n^2(\tau,w)d\tau^2+S^2(\tau,w)[d\chi^2+f^2_k(\chi)(d\theta^2+\sin^2\theta d\phi^2)]$$ 
we have $\epsilon=+1$ and
\be H^\tau_{~\tau}=-12L(\tau,w){\partial K^\chi_{~\chi}\over\partial w}+...\quad\hbox{with}\quad
L(\tau,w)\equiv -(K^\chi_{~\chi})^2+{\dot S^2+k n^2\over n^2S^2}\lbl{BH}\ee
and where $K^\chi_{~\chi}=-S'/S$, a prime denotes $\partial/\partial w$. Outside the membrane, spacetime is anti-de Sitter, $n$ and $S$ are given
in ref. [12], and $L(\tau,w)\to L_\pm$ (as an explicit calculation shows). In the vicinity and inside the 
membrane the function $L(\tau,w)$ is either continous with discontinuous
$w$-derivative (case of branes) or discontinous (case of shells) and can be modelled by the expressions 
(\ref{4.9})-(\ref{4.10}) in the text.


\end{document}